# Spoke formation in low temperature $E \times B$ plasmas. Transition from gradient-drift instability to ionization wave

J.P. Boeuf [a]

LAPLACE, Université de Toulouse, CNRS, INPT, UPS, 118 Route de Narbonne, 31062 Toulouse, France

**ABSTRACT**

Long wavelength plasma non-uniformities rotating in the azimuthal direction ("rotating spokes") have been observed in a number of experiments on Hall thrusters or magnetron discharges. We use a two-dimensional (2D), axial-azimuthal Particle-In-Cell Monte Carlo Collisions (PIC-MCC) model to study the formation of instabilities in a direct current (dc) magnetron discharge under conditions close to recent experiments. In spite of the simplified 2D geometry of the model, the simulations can reproduce the main features of the experimental results. At a given position above the cathode, corresponding to the spoke location, the simulations show large amplitude oscillations of the plasma density and a very sharp increase of the plasma potential and electron temperature at the leading edge of the spoke, as in time resolved probe measurements. Moreover, the simulations show that the instability evolves in time from a gradient-drift type of instability in the linear phase, to an ionization wave in the non-linear phase, with rotation in the $+E \times B$ direction in the first phase and in the $-E \times B$ direction in the second phase. The number of spokes is found to increase with pressure, as in experiments. The mechanisms of electron heating and the role of the $B \times \nabla B$ drift in electron heating and in the coherence and direction of spoke rotation are discussed.

## I. INTRODUCTION

Hall thrusters, magnetron discharges, or Penning discharges[1-8], are low temperature, partially magnetized plasma devices where electrons are strongly magnetized while ions are weakly magnetized. The electron Larmor radius is smaller than the characteristic dimensions of the plasma while the ion Larmor radius is not. An external magnetic field is placed perpendicular to the applied electric field or to the discharge current to increase the electron residence time in the device and allow ionization and plasma sustainment at low pressures (in the 1-50 mtorr range) and in relatively small devices (dimensions of a few cms). Electron confinement is efficient only if the $E \times B$ current is closed on itself, otherwise the confinement would be destroyed by the Hall effect. The geometry of these devices is therefore cylindrical, with $E \times B$ in the azimuthal direction, i.e. $E$ axial and $B$ radial or the opposite. Electrons are trapped by the magnetic field and are collisional while ions can be freely accelerated by the electric field. In regions with large electric fields, the electron drift velocity in the azimuthal, $E \times B$ direction, is E/B while ions are accelerated out of these regions before they can acquire the E/B azimuthal velocity (this is another way to say that these plasmas are partially magnetized).

The large difference between the azimuthal drift velocities of electrons and ions can generate charge separation which can lead to the development of instabilities. One of these instabilities develops when the electric field and the plasma density gradient are in the same direction[9] (gradient-drift or Simon-Hoh instability). This situation is very common in $E \times B$ plasma devices such as magnetron discharges[3, 4, 10] or in the channel of a Hall thruster[2]. Other types of instabilities can develop in regions where the electric field and plasma density gradients are in opposite direction, as in the exhaust region of a Hall thruster where particle models have shown the formation of an Electron Cyclotron Drift Instability[11-14].

In this paper we study, with the help of a 2D PIC-MCC model, the development of instabilities in a direct current (dc) magnetron discharge, under conditions similar to those of the experimental papers of Held et al.[15] and of Panjan et al.[16, 17]. The instabilities in magnetron discharges take the form of large-scale plasma non-uniformities rotating in the azimuthal direction and associated with rotating regions of enhanced light emission and ionization rate. These rotating ionization regions are generally called "spokes" and were first mentioned in the context of Hall thrusters by Janes and Lowder[18]. They have been more recently evidenced by high speed camera imaging in Hall thrusters[19-24] as well as in magnetron discharges[25-32]. An interesting feature of the spoke rotation discussed in several papers[17, 28, 31-33] is that the spoke can rotate in the $-E \times B$ direction in low power dc magnetrons and in the $+E \times B$ direction in high power pulsed magnetrons used in plasma processing applications (HiPIMS, High Power Impulse Magnetron Sputtering[5]).

---
[a] Electronic mail: jpb@laplace.univ-tlse.fr

Particle-In-Cell Monte Carlo Collisions (PIC-MCC) are now widely used to study the physics and instabilities in Hall thrusters[11-14, 34, 35], magnetron discharges[36-40], and Penning discharges[41-43]. Two-dimensional (2D) Particle-In-Cell Monte Carlo Collisions (PIC-MCC) simulations of magnetron discharges by Boeuf and Takahashi[38, 39] performed under conditions of dc magnetron discharges similar to those of the experiment of Ito et al.[31] were able to qualitatively reproduce the formation of spokes and the $-E \times B$ rotation.

In the experiments of Held al.[15] that are simulated in the present paper, a Langmuir probe has been used to measure the plasma density, plasma potential, electron temperature and electron energy probability function (EEPF) at a given location above the cathode, on the spoke's path, and as a function of time. The measurements show sharp increases of the plasma potential, electron density and plasma density when the spoke front reaches the probe, followed by a slower decay. The plasma properties present large amplitude oscillations as a function of time in the spoke region. The tail of the EEPF presents a sharp increase at the spoke front, followed by a quick decay. The large potential drop at the spoke front had been reported in previous experimental works on magnetron discharges[3, 17, 32].

The goal of the present paper is to complement the work of Refs.[38, 39] by more detailed (but still qualitative) comparisons with the experiments of Held et al.[15] and Panjan et al.[16, 17]. We compare the PIC-MCC simulations results with the probe measurements of Held et al., and the effect of pressure on the number of spokes with the experimental results of Panjan et al.. We also describe some interesting results of the particle simulations, which show that, starting from a 1D PIC-MCC stable solution of the dc magnetron discharge, the formation of the instability is consistent with the fluid theory and dispersion relation of gradient drift instabilities[9] in the linear phase but that the instability evolves into an ionization wave in the non-linear stage. The phase velocity of the instability changes direction from a $+E \times B$ direction in the linear phase, to a $-E \times B$ direction in the non-linear phase.

The PIC MCC model is briefly described in section II. The results are presented and compared to the experimental results in section III.

## II. PIC-MCC MODEL AND CONDITIONS OF THE SIMULATIONS

The PIC-MCC model is the same as in Refs.[38, 39].

The simulations are performed in a 2D Cartesian geometry in the plane defined by the axial and azimuthal directions of the magnetron (see Figure 1). The curvature of the magnetic field is not taken into account in this simplified geometry. The magnetic field is perpendicular to the simulation domain and is supposed to depend on the axial coordinate only. Periodic boundary conditions are used in the azimuthal direction.

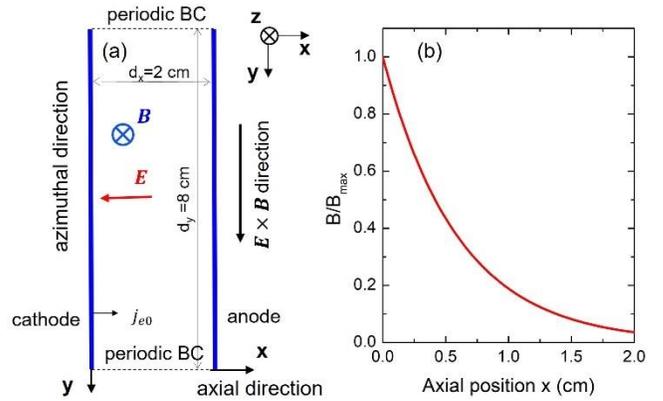

Figure 1: (a) Two-dimensional, axial-azimuthal simulation domain. The magnetic field is perpendicular to the simulation domain; (b) axial distribution of the perpendicular magnetic field of the form $B(x) = B_{max}\exp(-x/a)$, with $B_{max} = 120$ mT and $a = 0.3 \, d_x$ unless otherwise mentioned; $d_x = 2$ cm is the cathode-anode gap. $j_{e0}$ is the net current density of secondary electrons emitted by the cathode due to ion impact and is 0.03 A/m² in all the simulations (uniform along the cathode surface). The applied voltage is $V = 260$ V in most simulations. The positive $y$ direction is downward. The applied field $E$ is in the $-x$ direction, the magnetic field $B$ in the $+z$ direction and $E \times B$ is in the $+y$ direction.

The 1D version of this 2D PIC-MCC model (J-PIC code, see Ref.[44]) has also been used to study the 1D axial solutions of the same magnetron discharges, and to provide initial conditions for the 2D simulations.

The cathode-anode gap is $d_x = 2$ cm and the length of the simulation domain in the periodic azimuthal direction is $d_y = 8$ cm. A voltage of 260 V is applied between cathode and anode unless otherwise mentioned. The axial profile of the perpendicular magnetic field is given by:

$B(x) = B_{max}\exp(-x/a)$

with $B_{max} = 120$ mT and $a = 0.3 \, d_x$ in the simulations presented in section III.

Secondary emission due to ion impact is simulated by imposing a uniform net current density $j_{e0}$ of electrons emitted by the cathode, as in Ref.[39] ("net" current density means that each electron coming back to the cathode is re-emitted). The corresponding net secondary electron emission coefficient under ion impact, $\gamma_{net}$, can be deduced *a posteriori* (ratio of the ion current at the cathode to the net electron current emitted by the cathode). For a given, fixed current density $j_{e0}$ the secondary emission coefficient can therefore depend on the conditions (pressure, voltage magnetic field), in an unrealistic way. It is clear that care must be taken in the analysis of parameter studies performed with a constant $j_{e0}$ instead of a constant $\gamma_{net}$ but this method allows to limit the plasma density to values that can be handled by the particle model.

The simulations are performed in argon, at pressure between 0.6 Pa (close to the pressure in Held et al.[15]), and 2. Pa. As in Ref.[44], the electron–neutral cross sections used in the simulations are taken from the siglo database[45] of LXCat[46] (compilation of Phelps where the electron excitation levels are grouped in one single level), and the isotropic and backward scattering ion–neutral cross-sections for argon are taken from Phelps[47].

The number of (x,y) grid intervals is either (128×512) or (256×1024) depending on the plasma density (the grid spacing must satisfy $\delta x, \delta y < \lambda_{De}$, where $\lambda_{De}$ is the electron Debye length). The time step is such that $\delta t < 0.2/\omega_{pe}$ and must be sufficiently smaller than the minimum cyclotron rotation time around the magnetic field. $\delta t = 2 \times 10^{-11}$ s was used in most simulations. The number of particles per cell was between 30 and 70 in the 2D simulations presented in section III. OpenMP (Open Multi-Processing) parallelization with particle decomposition is used and the simulations presented here have been run on 12 or 24 threads.

To study the formation of instabilities, we first perform a 1D PIC-MCC simulation in the conditions above. The 1D solutions are stable for the conditions considered in this paper (small amplitude axial plasma oscillations can be present under some conditions, depending on the magnetic field profile, plasma density, etc…). We then use the 1D solution as an initial condition for the 2D simulation. This means that the initial axial positions and velocities of the charged particles in the 2D simulations are identical to those of the 1D steady state solution; the initial azimuthal positions are chosen randomly according to a uniform distribution. This allows the study of the linear development of the instability around the unperturbed, stable, steady state 1D solutions.

### III. SIMULATION RESULTS AND COMPARISONS WITH EXPERIMENTS

In the first part of this section (III.A) we show the results of 1D PIC-MCC simulations of magnetron discharges in the conditions described in section II. We then show (section III.B) the formation of instabilities in 2D axial-azimuthal simulations with initial conditions corresponding to the 1D stable solutions, and the transition to the spoke (ionization wave) regime. The linear phase of the instability is discussed in section III.C. The spoke regime is described in more detail in section III.D. Comparisons with probe measurements are presented in section III.E while considerations about spoke velocity, electron heating, and the effect of pressure are discussed in sections III.F, G, and H respectively.

#### A. 1D PIC-MCC simulations

Figure 2 shows the axial distributions, at steady state, of the charged particle densities, electric field, electron temperature and ionization rate obtained with the 1D PIC-MCC model in argon at a pressure of 0.6 Pa, cathode-anode gap $d_x$=2 cm, applied voltage $V$=260 V and net current density of electron emitted by at the cathode due to ion impact $j_{e0}$ = 0.03 A/m². The perpendicular magnetic field profile is given in Figure 2, with $B_{max}$ =120 mT. The total electron current density at the anode (close to the discharge current density) deduced from the simulation is $j_{ed}$ =5.1 A/m². This gives an electron multiplication $M = j_{ed}/j_{e0} = 170$ and a net secondary electron emission coefficient $\gamma_{net} = 1/(M-1) = 6 \times 10^{-3}$.

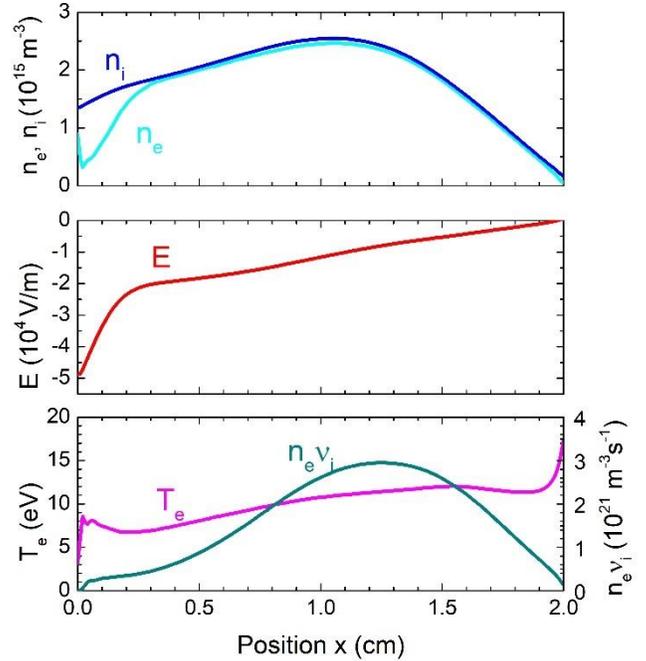

Figure 2 : Profiles of the electron density, ion density, electric field and ionization rate from the 1D PIC-MCC model in argon at a pressure $p$ =0.6 Pa and with a cathode-anode gap $d_x$=2 cm and an applied voltage $V$=260 V. The net electron density at the cathode is $j_{e0}$= 0.03 A/m². The axial profile of the perpendicular magnetic field is shown in Figure 1b and the maximum magnetic field at the cathode surface is $B_{max}$ =120 mT.

Assuming that the probability of reflection of an electron coming back to the cathode to be re-emitted is 0.1, a net secondary emission coefficient of $6 \times 10^{-3}$ corresponds to a "real" secondary electron emission by ion impact $\gamma = 6 \times 10^{-2}$, which is a reasonable order of magnitude for argon ions on a metal cathode[48].

We see on Figure 2 the cathode sheath and quasineutral plasma. Because of the large perpendicular magnetic field and low electron mobility there is no negative glow and the electric field is large in the quasineutral plasma. The electron temperature is large, on the order of 10 eV over the whole gap, and the location of maximum ionization rate is closer to the anode than to the cathode. The plasma density gradient on the anode side of the maximum plasma density is negative, i.e. of the same sign as the electric field. This suggests that gradient drift (Simon-Hoh) types of instabilities can form in this region if the azimuthal direction is taken into account.

#### B. 2D PIC-MCC simulations – Transition from gradient drift instability to ionization wave

As said above, the 2D simulations are started with initial conditions corresponding to the 1D stable solution described in the previous sub-section. This allows to study the formation of the 2D azimuthal instability around the stable solution, i.e. in the linear phase, and to follow the evolution from the linear stage to the non-linear, saturated stage.

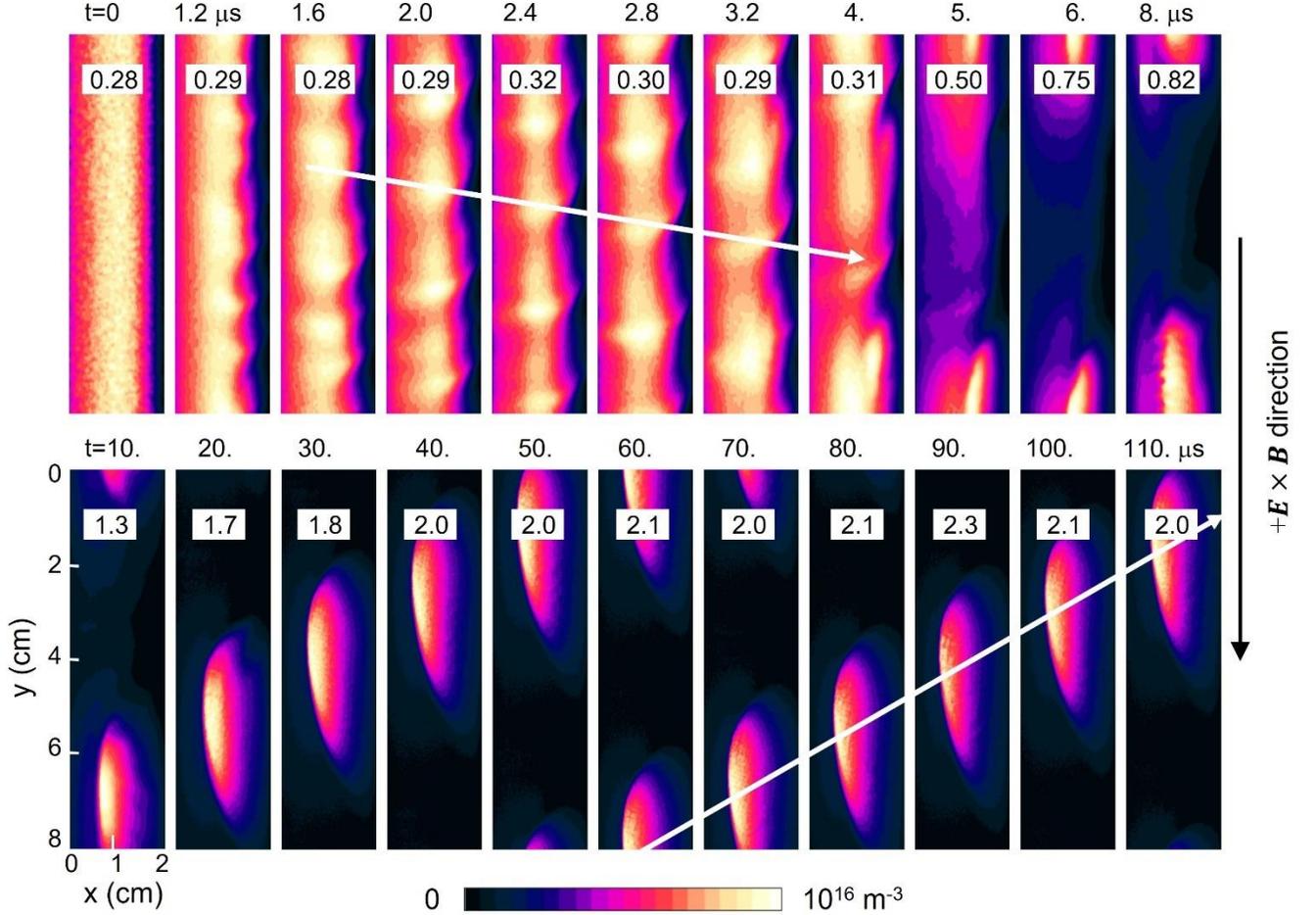

Figure 3: Color contour plots of the time evolution of the ion density (linear scale) from the 2D axial-azimuthal PIC-MCC simulation in the conditions of Figure 2 and with the magnetic profile of Figure 1b, $B_{max}$ =120 mT; argon, 0.6 Pa, 260 V dc, dimensions (2 cm × 8 cm), $j_{e0}$ = 0.03 A/m$^2$, (256×1024) grid intervals, time step $\delta t = 2 \times 10^{-11}$ s; initial conditions from the 1D simulation results of Figure 2, with charged particle uniformly distributed along the azimuthal direction. The time is indicated above each contour plot, and the unit (in 10$^{16}$ m$^{-3}$) is given in a white rectangle inside each plot.

Figure 3 shows such evolution in the conditions of Figure 2. We see that the instability forms in about 1 μs. The wavelength of the instability in the early stage of its development is slightly larger than 1 cm and increases in time until one single plasma non-uniformity is present in the azimuthal direction (at about 10 μs).

During the time interval [1-5] μs the non-uniformities move in the $+\mathbf{E}\times\mathbf{B}$ direction with a velocity around 8 km/s (the sound speed or argon ion for an electron temperature of 10 eV – see Figure 2, is close to 4.9 km/s). During this time the growth of the plasma density is not significant. After time t=4-5 μs a well-defined region of larger plasma density, that we will call "spoke" in the rest of this paper has formed and the growth of the plasma density due to ionization becomes significant till time t=40 μs. Between t=4 μs and t=8 μs the spoke velocity goes through zero and one can observe a reversal of the direction of the spoke velocity from the $+\mathbf{E}\times\mathbf{B}$ to the $-\mathbf{E}\times\mathbf{B}$ direction. After t=8 μs the spoke moves in the $-\mathbf{E}\times\mathbf{B}$ at a constant velocity slightly smaller than 1.5 km/s and without significant change in its shape and maximum plasma density. The motion of the spoke is very regular and coherent after that time. The simulation has been performed up to t=200 μs without any significant change in the spoke properties and velocity.

The spoke corresponds to the non-linear, saturated stage of the instability. We will see below that the non-linear stage of the instability after t=8 μs corresponds to an ionization wave. Before studying in detail the spoke properties in the ionization wave regime, we briefly discuss, in the next sub-section, the mechanisms responsible for the instability and the evolution of the instability from short wavelengths to larger wavelengths before the spoke formation.

### C. Linear stage of the instability – Dispersion relation

Fluid dispersion relations of low temperature partially magnetized $\mathbf{E}\times\mathbf{B}$ plasma are described in the article by Smolyakov et al.[9] (see also Refs. [43, 49, 50]). We consider purely azimuthal perturbations of the density and potential of the form $\exp(-i\omega t + iky)$, where $\omega$ is the complex wave frequency ($\omega = \omega_R + i\gamma$), $k$ is the wave number in the azimuthal direction. $\omega_R$, the real part of $\omega$ is the wave frequency and $\gamma$ is the growth rate (the perturbation grows when $\gamma > 0$).

Taking into account $\boldsymbol{E} \times \boldsymbol{B}$ drift, diamagnetic drift, magnetic field gradient, and including electron inertia, gyroviscosity, electron-neutral collisions and the effect of finite Debye length, the fluid dispersion relation can be written as[9, 43, 49] (assuming constant electron temperature):

$$k^2\lambda_{De}^2 + \frac{\omega_* - \omega_D + k^2\rho_e^2(\omega - \omega_0 + i\nu_{en})}{\omega - \omega_0 - \omega_D + k^2\rho_e^2(\omega - \omega_0 + i\nu_{en})} = \frac{k^2 c_s^2}{\omega^2} \quad (1)$$

where $k$ is the wave number in the azimuthal direction, $\rho_e = (eT_e/m)^{1/2}/\omega_{ce}$ is the electron Larmor radius ($\omega_{ce} = eB/m$, $m$ is the electron mass), $\omega_* = kv_*$, $v_* = -T_e/(BL_n)$ is the diamagnetic drift velocity and $L_n = [\partial(\ln n_0)/\partial x]^{-1}$, $\omega_0 = kv_0$, $v_0 = -E_0/B$ is the $\boldsymbol{E} \times \boldsymbol{B}$ drift velocity, $c_s = (eT_e/M)^{1/2}$ is the ion sound speed ($M$ is the ion mass) and $\nu_{en}$ is the electron-neutral collision frequency. $\omega_D$ is related to the magnetic field gradient: $\omega_D = kv_D$ with $v_D = 2T_e/(BL_B)$ and $L_B = [\partial(\ln B)/\partial x]^{-1}$. The index "zero" in $n_0$ and $E_0$ indicates equilibrium values (i.e. related to 1D results used as initial conditions). $n_0$ and $E_0$ depend on the axial position $x$). Note that with our notations (see Figure 1 and Figure 2), $\boldsymbol{E_0} = E_0\hat{\boldsymbol{x}}$, $\boldsymbol{B} = B\hat{\boldsymbol{z}}$, $\frac{\boldsymbol{E_0} \times \boldsymbol{B}}{B^2} = -\frac{E_0}{B}\hat{\boldsymbol{y}}$, so $E_0$ is negative, $B$ is positive, $L_B$ is negative, and $v_0$, $v_D$, $\omega_0$ and $\omega_D$ are positive; $L_n$ is negative and $v_*$ and $\omega_*$ are positive on the right side of the maximum plasma density in Figure 2, and change sign on the left side.

The classical collisionless Simon-Hoh instability is obtained in the limit of small values of $k\rho_e$ (i.e. long wavelength) and small Debye lengths. This gives:

$$\frac{\omega_* - \omega_D}{\omega - \omega_0} = \frac{k^2 c_s^2}{\omega^2} \quad (2)$$

For small wave frequencies, $\omega \ll \omega_0$, this equation has positive complex solutions, i.e. the instability is growing, when $(\omega_* - \omega_D)/\omega_0 > 0$. This corresponds to conditions where the electric field and the plasma density gradient are in the same direction and $\omega_* > \omega_D$. This is the case in the stable 1D solution of Figure 2, sufficiently close to the anode where we have $\frac{1}{|L_n|} > \frac{2}{|L_B|}$ ($|L_n|$ decreases close to the anode and $|L_B|$ is constant). Further away from the anode $|L_n|$ increases, $(\omega_* - \omega_D)$ becomes negative and the Simon-Hoh mode disappears (the magnetic field gradient stabilizes the Simon-Hoh mode).

To illustrate this, we can solve the dispersion relation (1) for the plasma parameters corresponding to a position close to the anode, $x = 1.8$ cm. The parameters corresponding to $x = 1.8$ cm can be obtained from the data of Figure 2: $n_0 = 0.8 \times 10^{15}$ m$^{-3}$, $E_0 = -2200$ V/m, $B = 6$ mT, $L_n = -2.2$ mm, $L_B = -6$ mm, $T_e = 11.4$ eV, $\rho_e = 1.35$ mm, $\lambda_{De} = 0.88$ mm, $c_s = 5.2$ km/s, $\nu_{en} = 2 \times 10^7 s^{-1}$.

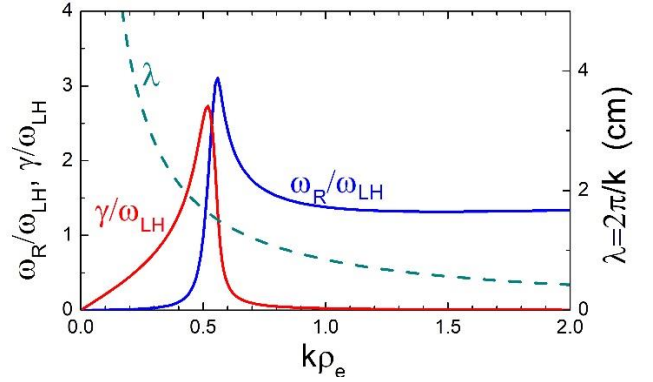

Figure 4: Angular frequency $\omega_R$ and growth rate $\gamma$ of the instability normalized to the lower hybrid frequency $\omega_{LH}$ calculated from the dispersion relation (1), calculate with the plasma and magnetic field parameters corresponding to the position $x = 1.8$ cm of Figure 2 ($\omega_{LH} \approx 4 \times 10^6$ rd/s). The dashed line represents the corresponding wavelength of the instability $\lambda = 2\pi/k$.

The solutions of the dispersion relation at this position, normalized to the lower hybrid frequency $\omega_{LH} = (eB/mM)^{1/2}$ are shown in Figure 4 ($\omega_{LH} \approx 4 \times 10^6$ rd/s at $x = 1.8$ cm). We see on Figure 4 that the instability grows for low values of $k\rho_e$ with a maximum growth rate $\gamma_{max}$ ($\gamma_{max} \approx 2.7\omega_{LH} \approx 11.\times 10^6$ s$^{-1}$), corresponding to a wavenumber $k_{\gamma_{max}} \approx 400$ m$^{-1}$ and to a wavelength $\lambda_{\gamma_{max}} = 2\pi/k_{\gamma_{max}}$ of about 1.6 cm, consistent with the first phase of Figure 3. At the maximum growth rate $\gamma_{max}$, the wave angular frequency $\omega_{R,\gamma_{max}}$ and the phase velocity $\omega_{R,\gamma_{max}}/k_{\gamma_{max}}$ are respectively on the order of $7 \times 10^6$ rd s$^{-1}$ and 17 km/s (note the sharp variations of $\omega_R$ for values of $k$ around $k_{\gamma_{max}}$ in Figure 4). The phase velocity estimated from Figure 3 was 8 km/s. The increase of the wavelength with time in Figure 3 is due to the deformation of the plasma density and electric field during the development of the instability.

As said above, the instability triggered at $x = 1.8$ cm corresponds to a Simon-Hoh mode that disappears further away from the anode because $|L_n|$ increases. $(\omega_* - \omega_D)$ goes through zero and changes sign at $x \lesssim 1.7$ cm. Around this point we observe, in the solutions of the dispersion relation (1), a transition from a Simon-Hoh mode to a lower-hybrid mode. For small values of $(\omega_* - \omega_D)$ the dispersion relation can be written[9], assuming small Debye length and $\omega, \omega_D \ll \omega_0$, neglecting the $k^2\rho_e^2$ term in the denominator of eq. (1), and noting that $c_s^2/\rho_e^2 = \omega_{LH}^2$:

$$\frac{(\omega_0 - i\nu_{en})}{\omega_0} = \frac{\omega_{LH}^2}{\omega^2} \quad (3)$$

For small $\nu_{en}$ with respect to $\omega_0$, this equation has solutions with positive growth rate[9]:

$$\omega = \omega_{LH}\left(1 + i\frac{\nu_{en}}{2\omega_0}\right) \quad (4)$$

This dispersion relation corresponds to a lower hybrid-mode ($\omega_R = \omega_{LH}$) destabilized by collisions and $\boldsymbol{E} \times \boldsymbol{B}$ flow.

Numerical solutions of the dispersion relation (1) for $x = 1.7$ cm (not shown here) confirm the transition from a Simon-Hoh or gradient-drift mode to a lower-hybrid mode, with maximum growth rates significantly smaller than in Figure 4.

In conclusion we can say that, in the conditions of Figure 3, the instability develops close to the anode surface through a Simon-Hoh or gradient-drift mode. The triggering of the instability is however strongly dependent on the relative values of the characteristic length of plasma density and magnetic field gradients. For values of the magnetic field gradient in the anode region smaller than that used in the conditions of Figure 3, the instability can be triggered by a lower-hybrid mode destabilized by collisions and $\boldsymbol{E} \times \boldsymbol{B}$ flow. We checked this numerically (not described here) by decreasing the magnetic field gradient in the anode region (keeping the same profile in the cathode region) and found that even in that case the instability evolves into an ionization wave similar to that of Figure 3, on a long enough time scale.

### D. Rotating spoke regime

We discuss here the steady state spoke regime that takes place after time t=10 μs in Figure 3. The electric potential, electron temperature and ionization rate at time t=80 μs are shown in Figure 5. The potential wave is correlated with the spoke motion shown on Figure 3. As found in Ref.[39], the spoke is associated with strong distortions of the electric potential, electron temperature and ionization rate.

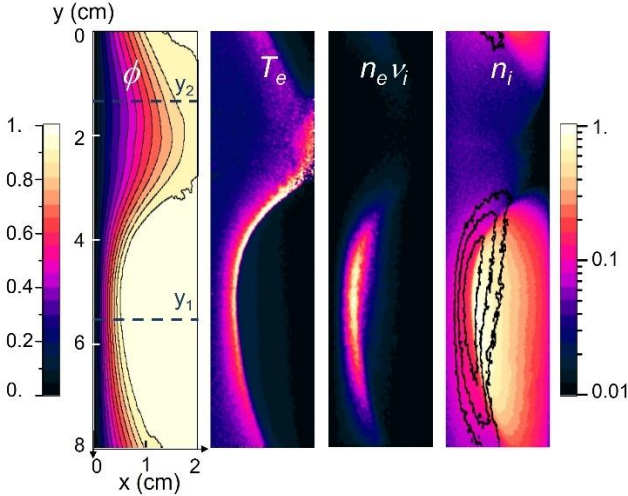

Figure 5 : 2D color contour plots of plasma potential, $\phi$, electron temperature, $T_e$, ionization rate, $n_e v_i$, and ion density, $n_i$, at time t=80 μs. in the conditions of Figure 3. $\phi$, $T_e$, and $n_e v_i$ are plotted on a linear scale (color bar on the left side), while $n_i$ is plotted on a two-decade log-scale (color bar on the right side). The potential contours (black lines) are plotted every 20 V from 0 to 260 V. The unit is 20 eV for the electron temperature, $1.4 \times 10^{22}$ m$^{-3}$s$^{-1}$ for the ionization rate, and $2 \times 10^{16}$ m$^{-3}$ for the ion density. Three contours of constant ionization rate (black lines) corresponding to 0.2, 0.4, and $0.8 \times 10^{22}$ m$^{-3}$s$^{-1}$ are superimposed on the ion density color contour plot. The axial profiles of the charged particle densities and electric field of Figure 6 are plotted at the azimuthal positions $y_1$ and $y_2$ indicated by the dashed lines on the potential plot.

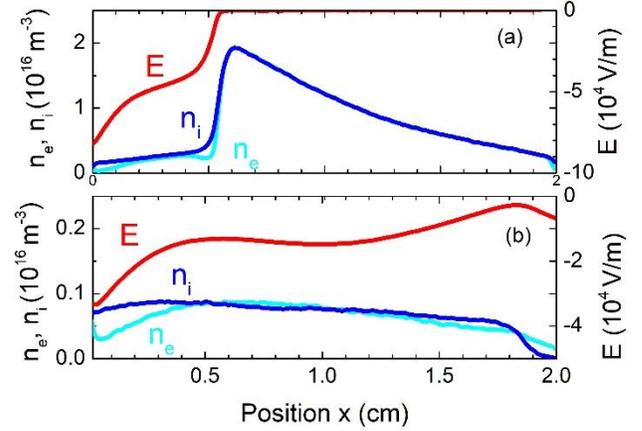

Figure 6 : Axial profiles of the charged particle densities, $n_e$, $n_i$, and electric field $E$, at two different azimuthal locations in the conditions of Figure 5 (time t=80 μs of Figure 3) corresponding to (a), $y_1$=5.56 cm, and (b), $y_2$=1.36 cm. These two azimuthal positions are indicated by dashed lines on the electric potential plot of Figure 5. Note that the scales of the plots in (a) and (b) are different.

The word "spoke", in the context of low temperature $\boldsymbol{E} \times \boldsymbol{B}$ plasmas, is generally used to refer to the region of intense light emission rotating azimuthally, which could be due to a local enhancement of the plasma density or electron temperature (and ionization and light emission) or both. In the magnetron discharge simulations of Figure 3 and Figure 5, it appears that the increase of the plasma density is the result of the local increase of the ionization rate and to the low electric field region behind the ionization region. In the rest of this paper we use the word "spoke" in a more general sense, to designate the ensemble formed by the ionization region (spoke front) plus the region of large plasma density generated behind the spoke front, and that is clearly visible in Figure 3 after time t=10 μs, and in Figure 5. The spoke is characterized by a large plasma density (about $2 \times 10^{16}$ m$^{-3}$ in the conditions of Figure 5) and a quasi-constant potential, close to the anode potential (electrons are trapped in this low electric field region, hence the high plasma density).

Figure 6 shows the axial profiles of the electric field and charged particle densities at two different azimuthal locations noted $y_1$ and $y_2$ on the plasma potential plot of Figure 5. One can see in Figure 6a the large electric field in the sheath between the cathode and the spoke at the azimuthal location $y_1$ and the very sharp increase of the plasma density at the spoke boundary. Outside the spoke in the azimuthal direction, the electric field penetrates in the quasineutral plasma as can be seen in the potential plot of Figure 5. The large electric field in the quasineutral plasma outside the spoke can also be seen in Figure 6b where the axial profiles of the charged particles densities and electric field are plotted at the location $y_2$.

The distribution of electron temperature shown in Figure 5 presents an important maximum at the boundary between the low electric field region in the spoke and the large field outside the spoke and is larger at the spoke front, i.e. on the upper side of the potential wave. The ionization rate is also maximum along the spoke boundary and is slightly shifted downward with respect to the electron temperature.

The maximum of ionization rate is located on the upper side of the maximum of plasma density (i.e. in the $-\boldsymbol{E} \times \boldsymbol{B}$ direction) as can be seen on the right plot of Figure 5 where the ion density is represented as color contours and the ionization rate as contour lines. This shows that the spoke motion is associated with an ionization wave moving in the $-\boldsymbol{E} \times \boldsymbol{B}$ direction. One can infer that the velocity of the spoke motion is related to the position of the ionization rate with respect to the maximum plasma density. The mechanisms of electron heating, which control in a large part the space distribution of the ionization rate, are therefore a determining factor in establishing the speed of the spoke rotation.

The conditions of the 1D simulation of Figure 2 and of the 2D simulation of Figure 3, Figure 5, and Figure 6 are identical (same current density of electrons emitted at the cathode, applied voltage, gas pressure and electrode gap). It is interesting to note that despite these identical conditions, the maximum plasma density is about ten times larger in the spoke of the 2D simulation than in the 1D simulation. Outside the spoke in the azimuthal direction, the 1D and 2D simulation results present similar values of the plasma density, and an important penetration of the axial electric field in the quasineutral plasma (compare Figure 2 and Figure 6b). The much larger plasma density and very small electric field in the spoke of the 2D simulation are due to the large ionization rate at the spoke front, which is responsible for the plasma generation behind the front. Note that despite the much larger maximum plasma density in the 2D result, the calculated total current density and the electron multiplication in the 2D simulation (respectively 4.5 A/m$^2$ and 150) are not much different from (and even slightly smaller than) those of the 1D simulation (respectively 5.1 A/m$^2$ and 170, see above).

Figure 7 shows the 2D distributions of the charge density $[n_i - n_e](x, y)$, of the electron current density $j_e(x, y)$ and current lines in the gap, and of the azimuthal profile $j_e(d_x, y)$ of the electron current density on the anode surface in the conditions of Figure 3, at time t=80 μs. The plot of the charge density shows that a double layer forms at the interface between the low electric field region in the spoke and the higher electric field around it. The ion sheath associated with this double layer is clearly visible on Figure 7, while the negative charge density is more diffuse in the spoke. The electron current density is maximum in the ion sheath region of the double layer. This corresponds to maximum values of the $\boldsymbol{E} \times \boldsymbol{B}$ and diamagnetic current density. Note that the electric field is perpendicular to the equipotential lines so the $\boldsymbol{E} \times \boldsymbol{B}$ current density follows the equipotential lines.

It appears on Figure 7a that the deformation of the plasma potential associated with the spoke creates a kind of short-circuit between cathode and anode and is responsible for the transport of a large electron current from the cathode region to the anode side. This can be seen on the electron current density profile on the anode surface, plotted in Figure 7c (note the different scales of the $j_e(x, y)$ and $j_e(d_x, y)$ plots).

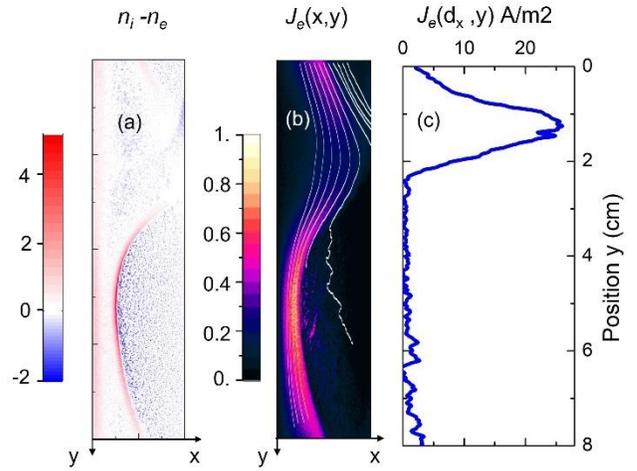

Figure 7: 2D distributions of (a) the charge density $n_i - n_e$, (units: 10$^{15}$ m$^{-3}$), and (b) the total electron current density $j_e$ (unit: 500 A/m$^2$); the colour contours show the intensity of the current density, the white lines show the current lines; (c) electron current density profile on the anode surface. Conditions of Figure 3 at time t=80 μs.

### E. Comparisons with probe measurements

Several experimental papers report probe measurements of the plasma properties in magnetron discharges. In this section we consider the recent measurements of Held et al.[15] who used Langmuir probes to measure the time evolution of plasma potential, electron density, electron temperature and electron energy probability functions (EEPF) in dc and pulsed magnetron discharges in the spoke region. The conditions of the measurements in the dc case (argon, 0.56 Pa, 255 V dc voltage) are close to those of the particle simulations described above. The magnetic field distribution used in the simulation (Figure 1b) is only a rough approximation of the magnetic field in the experiment, based on Figure 1 of Held et al.[15]. The Langmuir probe in the experiment was placed 6 mm above the cathode surface. The azimuthal period of the simulations (8 cm) is close to the length of the azimuthal line at the probe location in the experiment.

The probe measurements of Held et al. are shown in Figure 8. The measurements exhibit large amplitude oscillations of the electron density, electron temperature and plasma potential, with a very sharp increase of the plasma potential and electron temperature when the spoke reaches the probe. The modulation of the electron density is as large as 70 % and the plasma potential and electron temperature jumps at the spoke front are respectively on the order of 20 eV and 60 V.

The results of the 2D PIC-MCC simulations are displayed in Figure 9. An important difference between experiments and simulations that can be seen by comparing Figure 8 and Figure 9 is that the plasma density is about ten times smaller in the simulations. This is because the current density of electrons emitted at the cathode in the simulations has been imposed (0.03 A/m$^2$) in order to limit the maximum plasma density to numerically tractable values. Despite this difference, the time evolution of the plasma parameters in the simulation are qualitatively similar to those of the experiment.

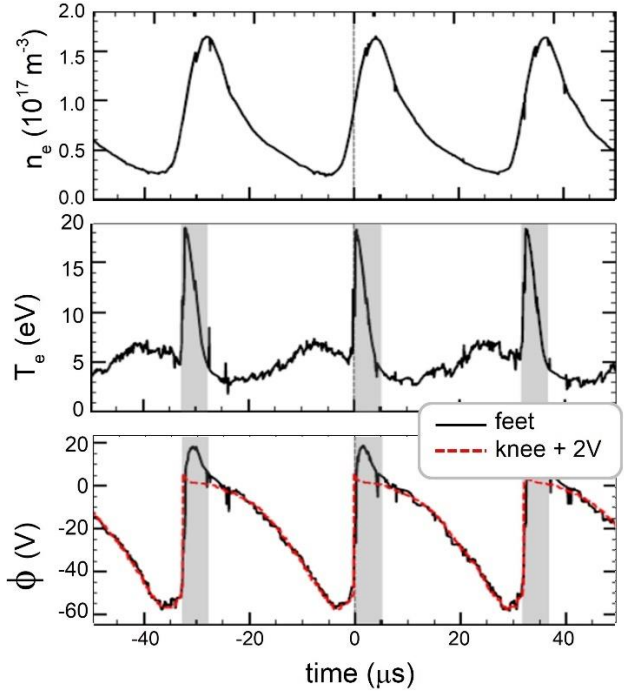 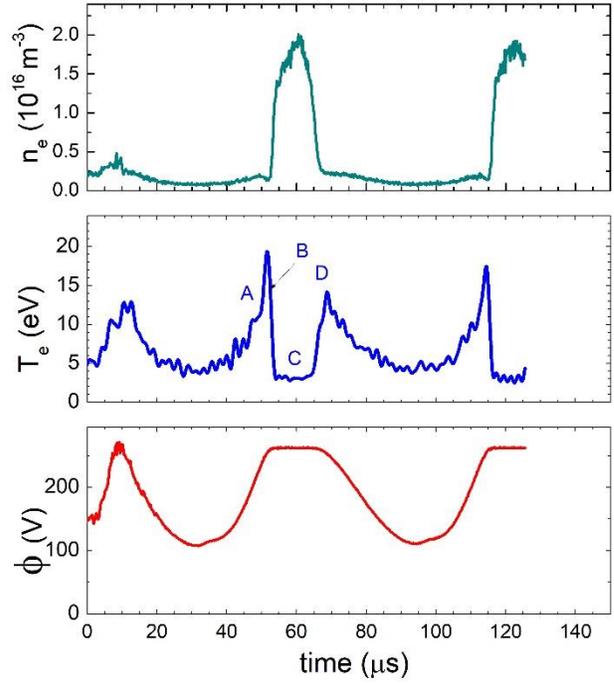

Figure 8 : Probe measurements by Held et al.[15] of the time evolution of electron density, $n_e$, electron temperature, $T_e$, and plasma potential $\phi$ in a dc magnetron discharge in argon at 0.56 Pa and with a dc voltage of 255 V. The probe is located 6 mm above the cathode surface on the spoke path. The gray areas indicate positions where the validity of the electron temperature and plasma potential are in question (see Held et al. for a complete description of these Langmuir probe measurements).

Figure 9: Time variations of electron density, $n_e$, electron temperature, $T_e$, and plasma potential $\phi$ from the 2D PIC-MCC simulation, 6 mm above the cathode surface ($x = 6$ mm), in the conditions of Figure 3 (argon, 0.6 Pa, dc voltage 260 V).

The simulation results of Figure 9 exhibit large amplitude oscillations of the plasma properties with abrupt increase of the electron temperature, plasma potential, and electron density when the spoke reaches the location of the numerical probe. The potential jump is larger in the simulation and, although the minimum and maximum electron temperature are similar in the experiment and simulation, the detailed time variations of the electron temperature present some discrepancy with those of the measurements.

For example, in the experiment, the electron temperature increases very sharply at the spoke front and decreases more slowly after the front. This is the opposite in the simulation results where the decay of electron temperature after the spoke (marked "B" in the electron temperature plot of Figure 9) is much faster than the temperature rise just at the spoke front ("A" in the temperature plot). One possible reason for the sharper electron temperature rise at the spoke front is the larger plasma density (and smaller Debye length, i.e. thinner double layer) in the experiment. The temperature after the spoke front quickly drops to about 3 eV in the simulation ("C" in the electron temperature plot of Figure 9), and increases again at the trailing edge of the spoke ("D" in the electron temperature plot). This is consistent with the 2D plots of the electron temperature in Figure 5).

The slower decay of the electron temperature in the experiments could be due to Coulomb collisions, which are not taken into account in the simulations.

It is interesting to look at the electron energy probability function (EEPF) in the region of the spoke front and to compare the EEPF deduced from the simulations, with the EEPF measured by Held et al. [15] (Figure 9 of Ref.[15]). Figure 10 shows the EEPF deduced from the PIC-MCC simulations at three different locations around the spoke front. The tail of the EEPF is strongly modulated in this region. The population of high energy electrons increases abruptly in the double layer region and decays in the high plasma density and low field region of the spoke. This is consistent with the heating of electrons flowing along the double layer in the spoke front. The residence time of electrons in the high-density plasma behind the spoke front is long because these electrons are trapped in this low electric field region. This explains the fast decay of the electron temperature in the simulations and the large population of low energy electrons in that region (Coulomb collisions could reduce this effect).

Finally, we note that the period of variations of the plasma parameters is shorter in the experiments of Held et al. (about 32 μs, see Figure 8) than in the simulation (60 μs, Figure 9). This indicates that the spoke velocity in the experiments (8 cm in 32 μs, i.e. 2.5 km/s) is about twice the velocity in the simulation (about 1.3 km/s according to Figure 9).

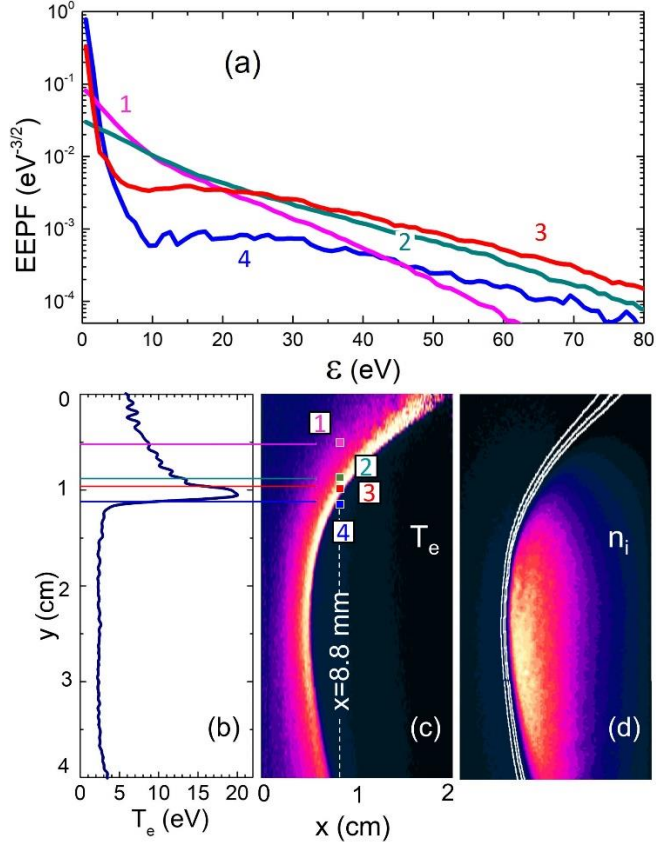

Figure 10: (a) Normalized electron energy probability function (EEPF) at four different azimuthal locations around the spoke front, 8.8 mm above the cathode surface ($x = 8.8$ mm) obtained from the PIC-MCC simulation at a given time in the conditions of Figure 3. The EEPF are integrated over squares of 80 μm side (indicated on (c)) in space, and over 10 ns in time; (b) azimuthal profile of the electron temperature at $x = 8.8$ mm in the conditions of (a) (showing only 4 cm of the 8 cm azimuthal length); (c) 2D distribution of electron temperature, and (d) of ion density. The azimuthal locations corresponding to the EEPF of (a) are indicated by colored squares and numbers on (c), and by lines on (b), with the same colors and numbers as the EEPF plots on (a). The three white lines on (d) correspond to contour of constant potential 240, 250, and 255 V.

#### F. Spoke velocity

The spoke velocity in dc magnetron discharge experiments in argon is in the $-\boldsymbol{E} \times \boldsymbol{B}$ direction, typically between 0 and 10 km/s, and is found to increase with increasing current or power (see, e.g., Figure 6 of Hecimovic and von Keudell[32]). In the high power, pulsed regime the spokes rotate in the $+\boldsymbol{E} \times \boldsymbol{B}$ direction with velocities up to 15 km/s[32].

We did not perform a systematic parameter study of the spoke velocity, but we noticed that the profile of the magnetic field has an influence on the spoke velocity in the simulation. For example in the simulations of Boeuf and Takahashi[38, 39] where the magnetic field was supposed to have a Gaussian shape (with a maximum of 100 mT at the cathode surface), the spoke velocity (in argon) was on the order of 10 km/s. In the present work we also performed simulations at 0.6 Pa with a magnetic field profile as in Refs.[38, 39] and a maximum magnetic field of 120 mT (same conditions as Figure 3 except for the magnetic field profile). The spoke velocity was about 4 km/s in that case (instead of 1.3 km/s for the case of Figure 3). Finally, the spoke velocity at a pressure of 2 Pa in the same conditions as Figure 3 (this case is described in section III.H below) was 1.5 km/s, slightly larger than the 0.6 Pa case of Figure 3.

We can conclude that the spoke velocity depends in a complex way on many parameters (pressure, magnetic field intensity and profile, power, etc…). This is because the spoke motion is due to an ionization wave and its velocity depends on the relative distributions of ionization rate and plasma density. In any case, we can expect that the order of magnitude of the spoke velocity is also related to the ion mass (through the ion velocity) since the motion of ions accelerated outside the spoke front is also a parameter controlling the spoke velocity. More work is necessary to better understand the parameters controlling the spoke velocity.

#### G. Electron heating and role of the $\boldsymbol{B} \times \boldsymbol{\nabla} B$ drift

As can be seen in Figure 5, the electron temperature and ionization rate are maximum in the region of the double layer around the spoke. Electron heating is therefore larger along the double layer. We showed in Refs.[38, 39] that the electron drift due to the magnetic field gradient ("$\nabla B$ drift") can play an important role in electron heating in magnetron discharges. The $\boldsymbol{\nabla} B$ electron drift velocity is given by[51] $\boldsymbol{v}_{\nabla B} = -1/2\, \rho_e v_\perp \boldsymbol{B} \times \boldsymbol{\nabla} B/B^2 = -\varepsilon_\perp \boldsymbol{B} \times \boldsymbol{\nabla} B/B^3$ where $v_\perp$ is the electron velocity and $\varepsilon_\perp$ the electron energy in eV, perpendicular to the magnetic field. In the conditions of the simulations considered here (see Figure 1) $\boldsymbol{v}_{\nabla B}$ is directed in the azimuthal direction, in the positive $y$ direction.

The electron heating or cooling per electron due to the $\boldsymbol{\nabla} B$ drift is given by:

$$\theta_{\boldsymbol{v}_{\nabla B}} = -\boldsymbol{v}_{\nabla B} \cdot \boldsymbol{E} = \varepsilon_\perp E_y / (B L_B) \qquad (5)$$

where $L_B = B/\partial_x B$. Considering the direction of the azimuthal electric field $E_y$ around the spoke that can be deduced from the equipotential contours of Figure 5, we can conclude that the $\nabla B$ drift contributes to electron heating on the upper side of the potential wave ($E_y$ and $L_B$ are negative in eq. (5)) and to electron cooling on the lower side ($E_y$ positive, $L_B$ negative). Note that in the absence of collisions, electron heating and cooling would cancel each-other because of the adiabaticity of the electron trajectories. The fact that there is a net heating in our conditions is due to collisions: electrons can lose energy through inelastic collisions before $\nabla B$ drift cooling becomes important.

Non-adiabatic trajectories leading to electron heating is possible when sharp spatial variations of the electric field such that $|\nabla E|/B > \omega_{ce}$ are present[52]. This situation occurs in collisionless shocks in space plasmas[53-55]. We checked that the condition $|\nabla E|/B > \omega_{ce}$ is not satisfied in our simulations where the plasma density is limited because of numerical constraints.

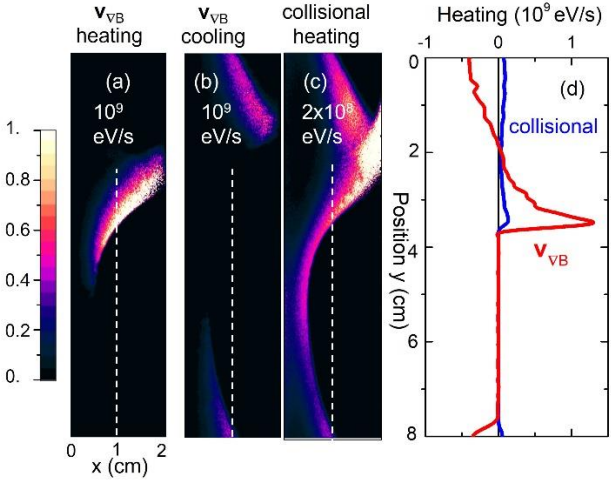

Figure 11 : Comparisons of the 2D distributions of the heating per electron due to, (a), (b) $\nabla B$ drift, $\theta_{v_{\nabla B}}$, eq. (5), with positive contribution in (a) and negative contribution in (b), and (c), to collisional heating, $\theta_E$, eq. (6); (d), azimuthal profiles at $x = 1$ cm (dashed line of the 2D contour plots (a), (b), and (c)) of the $\nabla B$ drift and collisional electron heating. Note the different scales, $10^9$ in (a), (b), and $2 \times 10^8$ eV/s in (c).

This condition could however be reached under the much higher plasma densities and smaller Debye lengths (i.e. thinner double layer) of high power magnetrons discharges, leading to larger values of $|\nabla E|/B$ in the double layer. This question is left for further studies.

Since the electric field is large in the double layer around the spoke, another possible contribution to electron heating is the classical, collisional heating. Neglecting the diamagnetic term, the contribution of collisional heating to the total electron heating can be written as:

$$\theta_E = -v_{\parallel E} \cdot E = \frac{\nu}{\Omega_{ce}} \frac{E^2}{B} \qquad (6)$$

where $v_{\parallel E}$ is the electron drift velocity parallel to the electric field.

The contribution of the $v_{\nabla B}$ and $v_{\parallel E}$ (collisional) heating to the total electron heating in the conditions of Figure 5 are shown on Figure 11. Figure 11a and Figure 11b confirm that the $\nabla B$ drift contributes to electron heating on one side of the potential wave and to electron cooling on the other side. Collisional electron heating is present all along the double layer, but its magnitude is significantly smaller than the heating due to $\nabla B$ drift. This is also true in the larger pressure condition (2 Pa) of section III.H. This confirms that the $\nabla B$ drift plays an essential role in the spoke formation and sustainment in low power dc magnetron discharges. Electron heating on the upper side of the potential wave due to $\nabla B$ drift and electron cooling on the lower side tend to limit the ionization region to the upper side, i.e. above the region of large plasma density. This clearly explains the rotation of the spoke in the $-E \times B$ direction and the coherence of the spoke rotation.

To confirm this conclusion, we performed simulations with uniform magnetic fields, in order to remove the $\nabla B$ drift contribution to electron heating. In a previous work[36, 37] we also performed 2D PIC-MCC simulations of magnetron discharges under uniform magnetic field, but this was done in cylindrical geometry, i.e. with axial magnetic field and cylindrical cathode and anode. In the cylindrical configuration as well as in the planar magnetron configuration (the present work) with uniform magnetic field, the particle simulation predicts the rotation of plasma non-uniformities in the $+E \times B$ direction. These non-uniformities are relatively coherent or much more chaotic at steady state, depending on the conditions, but they are never as regular and coherent as the spoke in a magnetic field gradient described in the present work and in Refs.[38, 39]. Moreover, in the simulations in a uniform magnetic field, we never found conditions where most of the ionization took place in the spoke (ionization in the cathode sheath was always dominant). For example, one can see in Figure 9 of Ref. [37] or in Figure 2 of Ref. [36] that ionization in the spoke is a very small fraction of the total ionization.

### H. Influence of pressure

We have seen above that electron heating is larger in the ion sheath region of the double layer around the spoke, on the upper side of the potential wave, due to a combination of $\nabla B$ electron drift and (to a lesser extent) collisional cross-field transport. Electrons heated in this region continue flowing in the $E \times B$ direction along the ion sheath around the spoke. They lose energy along this line due to inelastic collisions and of $\nabla B$ drift cooling. The mean free path for ionization in the $E \times B$ direction can be estimated as $\lambda_i = v_d/\nu_i$ where $v_d = E/B$ is the $E \times B$ drift velocity and $\nu_i$ the ionization frequency. The simulation in the conditions of Figure 3 gives a maximum of $\nu_i$ on the order of $10^7$ s$^{-1}$ in the electron heating region and a drift velocity $v_d$ between $5 \times 10^5$ and $10^6$ m/s. This gives a value of the ionization mean free path between 5 and 10 cm which is on the order of the azimuthal length in the conditions of Figure 3. It seems reasonable to expect an increase of the number of spokes as the ionization mean free path decreases, i.e. as the pressure increases, as discussed by Panjan et al.[16].

This is confirmed by the particle model. PIC-MCC simulations performed at a pressure of 2 Pa and applied voltage $V$=280 V (other conditions being the same as in Figure 3) predict the formation of two spokes in the azimuthal direction, as can be seen in Figure 12. At 2 Pa, we can expect an ionization mean free path smaller than at 0.6 Pa by a factor of about 3. This is consistent with the length of the ionization region in Figure 12 (2 Pa) compared with Figure 3 (0.6 Pa).

The consequence is that the number of spokes increases with gas pressure. This has been found in experiments as discussed in detail in the article by Panjan et al.[16].

The electron current density at the anode in these conditions (2 Pa, 280 V) is $j_{ed}$ =7.25 A/m$^2$, slightly larger than at 0.6 Pa, 5.1 A/m$^2$. This corresponds to an electron multiplication of 242 and a net secondary electron emission coefficient of $4 \times 10^{-3}$.

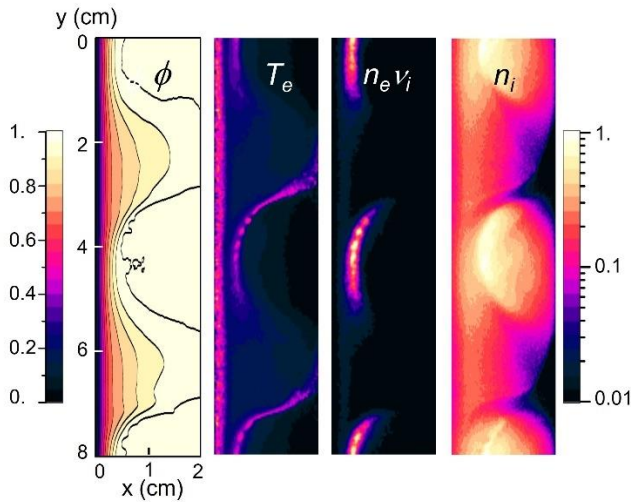

Figure 12: 2D color contour plots of plasma potential, $\phi$, electron temperature, $T_e$, ionization rate, $n_e \nu_i$, and ion density, $n_i$ in argon at a pressure $p = 2$ Pa and an applied voltage $V = 280$ V, all the other conditions being the same as in Figure 3. $\phi$, $T_e$, and $n_e \nu_i$ are plotted on a linear scale (color bar on the left side) while $n_i$ is plotted on a two-decade log-scale (color bar on the right side). The potential contours are plotted every 20 V from 0 to 280 V. The unit is 20 eV for the electron temperature, $3 \times 10^{22}$ m$^{-3}$s$^{-1}$ for the ionization rate, and $1.5 \times 10^{16}$ m$^{-3}$ for the ion density. The spokes move in the $-\boldsymbol{E} \times \boldsymbol{B}$ direction, as in the 0.6 Pa case.

If the pressure is further increased, the number of spokes continues to increase. The spokes become closer to each other, and the modulation of the plasma density decreases. We performed simulations at 3.3 Pa, other conditions being the same as in Figure 3. At this pressure the spokes had partially merged, the modulation of the plasma parameters were much less important and less coherent than in Figure 9 and the plasma non-uniformities were rotating in the $+\boldsymbol{E} \times \boldsymbol{B}$ direction.

## CONCLUSION

Two-dimensional PIC-MCC simulations of a dc magnetron under typical conditions (argon, pressure between 0.6 and 2 Pa, magnetic field on the cathode surface of 120 mT decreasing over 2 cm) have been performed to study the formation and properties of rotating spokes. These simulations confirm and complement previously published results[38, 39], and are qualitatively compared with recent experiments. The work described in the present paper can be summarized as follows.

- The results confirm that the magnetic field gradient plays an important role in electron heating and in the structuration and coherence of rotating spokes in magnetron discharges.
- The spoke is associated with a deformation of the plasma potential with equipotential lines oscillating between cathode and anode.
- A double layer forms along the equipotential line separating a region of low electric field and high plasma density (the spoke) and a region of larger electric field and low plasma density.
- A large $\boldsymbol{E} \times \boldsymbol{B}$ electron current flows along and around this line, and this contributes to electron transport across the magnetic field from the cathode side to the anode region.
- Electron heating takes place in the double layer, with a dominant contribution of $\boldsymbol{\nabla} B$ drift. Collisional heating is present in the double layer but is significantly lower than $\boldsymbol{\nabla} B$ drift heating.
- $\boldsymbol{\nabla} B$ drift heating followed by intense ionization takes place on the part of the double layer that is "above" (i.e. in the $-\boldsymbol{E} \times \boldsymbol{B}$ direction with respect to) the region of large plasma density, while $\boldsymbol{\nabla} B$ drift cooling takes place on the other side. This explains the $-\boldsymbol{E} \times \boldsymbol{B}$ rotation of the spoke: the spoke is an ionization wave. The well-defined location of the ionization region due to $\boldsymbol{\nabla} B$ drift heating is responsible for the coherence of the spoke rotation.
- By starting the 2D simulations with initial conditions corresponding to 1D stable steady state results, one can study the linear and non-linear stages of the instability. We find that gradient drift instabilities dominate in the linear stage while the instability evolves into an ionization wave associated with a spoke in the non-linear stage. The rotation of the instability therefore starts in the $+\boldsymbol{E} \times \boldsymbol{B}$ direction during the linear stage and changes to $-\boldsymbol{E} \times \boldsymbol{B}$ in the non-linear stage.
- The good qualitative agreement between experimental probe measurements[15] and simulations tends to confirm the understanding of the spoke regime provided by the model.
- The increase of the number of spokes with pressure observed in the experiments and attributed to a decrease in the electron mean free path[16] is also found in the simulations.
- The spoke velocity, although certainly related to the ion mass is a complex function of several parameters including magnetic field intensity and profile, discharge power, pressure. More work is needed to understand this dependence.

## DATA AVAILABILITY

The data that support the findings of this study are available from the corresponding author upon reasonable request.

## ACKNOWLEDGEMENTS

The author would like to acknowledge enlightening discussions with A. Smolyakov on instabilities in $\boldsymbol{E} \times \boldsymbol{B}$ plasmas and associated dispersion relations.

## REFERENCES

1. D. M. Goebel and I. Katz, *Fundamentals of Electric Propulsion: Ion and Hall Thrusters*. (Wiley, 2008).
2. J.-P. Boeuf, *Tutorial: Physics and modeling of Hall thrusters*, J. Appl. Phys. **121**, 011101 (2017).
3. A. Anders, *Tutorial: Reactive high power impulse magnetron sputtering (R-HiPIMS)*, J. Appl. Phys. **121**, 171101 (2017).
4. J. T. Gudmundsson, *Physics and technology of magnetron sputtering discharges*, Plasma Sources Sci. Technol. **29**, 113001 (2020).

5. J. T. Gudmundsson, N. Brenning, D. Lundin and U. Helmersson, *High power impulse magnetron sputtering discharge*, J. Vac. Sci. Technol. A **30**, 030801 (2012).
6. E. B. Hooper, *A Review of Reflex and Penning Discharges*, Adv. Electronics Electron Phys. **27**, 295 (1969).
7. S. N. Abolmasov, *Physics and engineering of crossed-field discharge devices*, Plasma Sources Sci. Technol. **21**, 035006 (2012).
8. I. D. Kaganovich, A. Smolyakov, Y. Raitses, E. Ahedo, I. G. Mikellides, B. Jorns, F. Taccogna, R. Gueroult, S. Tsikata, A. Bourdon, J.-P. Boeuf, M. Keidar, A. T. Powis, M. Merino, M. Cappelli, K. Hara, J. A. Carlsson, N. J. Fisch, P. Chabert, I. Schweigert, T. Lafleur, K. Matyash, A. V. Khrabrov, R. W. Boswell and A. Fruchtman, *Physics of E × B discharges relevant to plasma propulsion and similar technologies*, Phys. Plasmas **27**, 120601 (2020).
9. A. I. Smolyakov, O. Chapurin, W. Frias, O. Koshkarov, I. Romadanov, T. Tang, M. Umansky, Y. Raitses, I. D. Kaganovich and V. P. Lakhin, *Fluid theory and simulations of instabilities, turbulent transport and coherent structures in partially-magnetized plasmas of E × B discharges*, Plasma Phys. Controlled Fusion **59**, 014041 (2017).
10. M. Rudolph, N. Brenning, H. Hajihoseini, M. A. Raadu, T. M. Minea, A. Anders, J. T. Gudmundsson and D. Lundin, *Influence of the magnetic field on the discharge physics of a high power impulse magnetron sputtering discharge*, Journal of Physics D: Applied Physics **55**, 015202 (2021).
11. J. C. Adam, A. Héron and G. Laval, *Study of stationary plasma thrusters using two-dimensional fully kinetic simulations*, Phys. Plasmas **11**, 295 (2004).
12. J. P. Boeuf and L. Garrigues, *E x B electron drift instability in Hall thrusters: Particle-in-cell simulations vs. theory*, Phys. Plasmas **25** (2018).
13. T. Charoy, J. P. Boeuf, A. Bourdon, J. A. Carlsson, P. Chabert, B. Cuenot, D. Eremin, L. Garrigues, K. Hara, I. D. Kaganovich, A. T. Powis, A. Smolyakov, D. Sydorenko, A. Tavant, O. Vermorel and W. Villafana, *2D axial-azimuthal particle-in-cell benchmark for low-temperature partially magnetized plasmas*, Plasma Sources Sci. Technol. **28**, 105010 (2019).
14. A. Smolyakov, T. Zintel, L. Couedel, D. Sydorenko, A. Umnov, E. Sorokina and N. Marusov, *Anomalous Electron Transport in One-Dimensional Electron Cyclotron Drift Turbulence*, Plasma Physics Reports **46**, 496 (2020).
15. J. Held, M. George and A. von Keudell, *Spoke-resolved electron density, temperature and potential in direct current magnetron sputtering and HiPIMS discharges*, Plasma Sources Sci. Technol **31**, 085013 (2022).
16. M. Panjan, S. Loquai, J. E. Klemberg-Sapieha and L. Martinu, *Non-uniform plasma distribution in dc magnetron sputtering: origin, shape and structuring of spokes*, Plasma Sources Sci. and Technol. **24**, 065010 (2015).
17. M. Panjan and A. Anders, *Plasma potential of a moving ionization zone in DC magnetron sputtering*, J. Appl. Phys. **121**, 063302 (2017).
18. C. S. Janes and R. S. Lowder, *Anomalous Electron Diffusion and Ion Acceleration in a LowDensity Plasma*, Phys. Fluids **9**, 1115 (1966).
19. J. B. Parker, Y. Raitses and N. J. Fisch, *Transition in electron transport in a cylindrical Hall thruster*, Appl. Phys. Lett. **97**, 091501 (2010).
20. C. L. Ellison, Y. Raitses and N. J. Fisch, *Cross-field electron transport induced by rotating spoke in a cylindrical Hall thruster*, Phys. Plasmas **19**, 013503 (2012).
21. M. McDonald and A. D. Gallimore, *Rotating Spoke Instabilities in Hall Thrusters*, IEEE Trans. Plasma Sci. **39**, 2952 (2011).
22. M. J. Sekerak, A. D. Gallimore, D. L. Brown, R. R. Hofer and J. E. Polk, *Mode Transitions in Hall-Effect Thrusters Induced by Variable Magnetic Field Strength*, Journal of Propulsion and Power **32**, 903 (2016).
23. S. Mazouffre, L. Grimaud, S. Tsikata, K. Matyash and R. Schneider, *Rotating spoke instabilities in a wall-less Hall thruster: experiments*, Plasma Sources Sci. and Technol. **28**, 054002 (2019).
24. A. Guglielmi, F. Gaboriau and J. P. Boeuf, *Simultaneous measurements of axial motion and azimuthal rotation of non-uniformities ("spokes") in a Hall thruster*, Phys. Plasmas **29**, 112108 (2022).
25. A. Anders, *Self-organization and self-limitation in high power impulse magnetron sputtering*, Appl. Phys. Lett. **100**, 224104 (2012).
26. A. Anders, P. Ni and A. Rauch, *Drifting localization of ionization runaway: Unraveling the nature of anomalous transport in high power impulse magnetron sputtering*, J. Appl. Phys. **111**, 053304 (2012).
27. J. Winter, A. Hecimovic, T. de los Arcos, M. Böke and V. Schulz-von der Gathen, *Instabilities in high-power impulse magnetron plasmas: from stochasticity to periodicity*, J. Phys. D: Applied Physics **46**, 084007 (2013).
28. Y. Yang, J. Liu, L. Liu and A. Anders, *Propagation direction reversal of ionization zones in the transition between high and low current magnetron sputtering*, Appl. Phys. Lett. **105**, 254101 (2014).
29. A. Hecimovic, M. Böke and J. Winter, *The characteristic shape of emission profiles of plasma spokes in HiPIMS: the role of secondary electrons*, J. Phys. D: Applied Physics **47**, 102003 (2014).
30. A. Hecimovic, V. Schulz-von der Gathen, M. Böke, A. von Keudell and J. Winter, *Spoke transitions in HiPIMS discharges*, Plasma Sources Sci. and Technol. **24**, 045005 (2015).
31. T. Ito, C. V. Young and M. A. Cappelli, *Self-organization in planar magnetron microdischarge plasmas*, Appl. Phys. Lett. **106**, 254104 (2015).
32. A. Hecimovic and A. von Keudell, *Spokes in high power impulse magnetron sputtering plasmas*, J. Phys. D: Applied Physics **51**, 453001 (2018).
33. A. Hecimovic, C. Maszl, V. Schulz-von der Gathen, M. Böke and A. von Keudell, *Spoke rotation reversal in magnetron discharges of aluminium, chromium and titanium*, Plasma Sources Sci. and Technol. **25**, 035001 (2016).
34. T. Lafleur, S. D. Baalrud and P. Chabert, *Theory for the anomalous electron transport in Hall effect thrusters. I. Insights from particle-in-cell simulations*, Phys. Plasmas **23**, 053502 (2016).
35. W. Villafana, F. Petronio, A. C. Denig, M. J. Jimenez, D. Eremin, L. Garrigues, F. Taccogna, A. Alvarez-Laguna, J. P. Boeuf, A. Bourdon, P. Chabert, T. Charoy, B. Cuenot, K. Hara, F. Pechereau, A. Smolyakov, D. Sydorenko, A. Tavant and O. Vermorel, *2D radial-azimuthal particle-in-cell benchmark for E × B discharges*, Plasma Sources Sci. Technol. **30**, 075002 (2021).

36. J.-P. Boeuf and B. Chaudhury, *Rotating Instability in Low-Temperature Magnetized Plasmas*, Phys. Rev. Lett. **111** (2013).
37. J.-P. Boeuf, *Rotating structures in low temperature magnetized plasmas; insight from particle simulations*, Front. Phys. **2**, 74 (2014).
38. J. P. Boeuf and M. Takahashi, *Rotating Spokes, Ionization Instability, and Electron Vortices in Partially Magnetized E × B Plasmas*, Phys. Rev. Lett. **124**, 185005 (2020).
39. J. P. Boeuf and M. Takahashi, *New insights into the physics of rotating spokes in partially magnetized E × B plasmas*, Phys. Plasmas **27**, 083520 (2020).
40. M. Sengupta, A. Smolyakov and Y. Raitses, *Restructuring of rotating spokes in response to changes in the radial electric field and the neutral pressure of a cylindrical magnetron plasma*, J. Appl. Phys. **129**, 223302 (2021).
41. A. T. Powis, J. A. Carlsson, I. Kaganovich, Y. Raitses and A. Smolyakov, *Scaling of spoke rotation frequency within a Penning discharge*, Phys. Plasmas **25**, 072110 (2018).
42. J. Carlsson, I. Kaganovich, A. Powis, Y. Raitses, I. Romadanov and A. Smolyakov, *Particle-in-cell simulations of anomalous transport in a Penning discharge*, Phys. Plasmas **25**, 061201 (2018).
43. M. Tyushev, M. Zadeh, V. Sharma, M. Sengupta, Y. Raitses, J. P. Boeuf and A. Smolyakov, *Azimuthal structures and turbulent transport in Penning discharges*, Submitted Physics of Plasmas (2022).
44. J. P. Boeuf, *Ionization waves (striations) in a low-current plasma column revisited with kinetic and fluid models*, Phys. Plasmas **29**, 022105 (2022).
45. SigloDatabase, www.lxcat.net, retrieved on November 3, 2021.
46. E. Carbone, W. Graef, G. Hagelaar, D. Boer, M. M. Hopkins, J. C. Stephens, B. T. Yee, S. Pancheshnyi, J. van Dijk and L. Pitchford, *Data Needs for Modeling Low-Temperature Non-Equilibrium Plasmas: The LXCat Project, History, Perspectives and a Tutorial*, Atoms **9**, 16 (2021).
47. A. V. Phelps, *The application of scattering cross sections to ion flux models in discharge sheaths*, J. Appl. Phys. **76**, 747 (1994).
48. A. V. Phelps and Z. L. Petrović, *Cold-cathode discharges and breakdown in argon: surface and gas phase production of secondary electrons*, Plasma Sources Sci. Technol. **8**, R21 (1999).
49. W. Frias, A. I. Smolyakov, I. D. Kaganovich and Y. Raitses, *Long wavelength gradient drift instability in Hall plasma devices. I. Fluid theory*, Phys. Plasmas **19**, 072112 (2012).
50. L. Xu, D. Eremin, A. Smolyakov, D. Krüger, K. Köhn and R. P. Brinkmann, *Formation mechanism of the rotating spoke in partially magnetized plasmas*, Plasma Sources Sci. Technol, submitted (2022).
51. F. F. Chen, *Introduction to Plasma Physics and Controlled Fusion*. (Plenum Press, New-York, 1984).
52. K. Cole, *Effects of crossed magnetic and (spatially dependent) electric fields on charged particle motion*, Planet. Space Sci. **24**, 515 (1976).
53. M. Balikhin, M. Gedalin and A. Petrukovich, *New mechanism for electron heating in shocks*, Phys. Rev. Lett. **70**, 1259 (1993).
54. B. Lembège, P. Savoini, M. Balikhin, S. Walker and V. Krasnoselskikh, *Demagnetization of transmitted electrons through a quasi-perpendicular collisionless shock*, J. Geophys. Res.-Space **108**, 1256 (2003).
55. V. See, R. F. Cameron and S. J. Schwartz, *Non-adiabatic electron behaviour due to short-scale electric field structures at collisionless shock waves*, Ann. Geophys. **31**, 639 (2013).